\documentclass[a4paper,12pt]{article}
\usepackage[pctex32]{graphics}
\usepackage{amssymb,amsmath}
\usepackage{amsmath,amssymb}
\usepackage{latexsym}
\usepackage{epsfig}
\usepackage[english]{babel}

\newcommand{\be}{\begin{equation}}
\newcommand{\ee}{\end{equation}}
\newcommand{\ba}{\begin{eqnarray}}
\newcommand{\ea}{\end{eqnarray}}

\begin{document}

\begin{titlepage}

\vspace{5mm}

\begin{center}

{\Large \bf Entropy of black hole \\ in the infinite-derivative gravity}

\vskip .6cm

\centerline{\large
 Yun Soo Myung$^{a}$}

\vskip .6cm

{Institute of Basic Science and Department of Computer Simulation,
\\Inje University, Gimhae 50834, Korea \\}

\end{center}

\begin{center}
\underline{Abstract}
\end{center}
We compute the Wald  entropy of the Schwarzschild black hole in the ghost-free, infinite derivative gravity that is quadratic in curvature.
This is not  given purely  by the area law but includes  an additional contribution depending on the power of d'Alembertian operator,
when requiring that the massless graviton be the only propagating mode in the Minkowski sapcetime.

 \vskip .6cm

 \\

\vskip 0.8cm

\vspace{15pt} \baselineskip=18pt

\noindent $^a$ysmyung@inje.ac.kr \\

\thispagestyle{empty}
\end{titlepage}

\newpage
\section{Introduction}
It was shown recently  that the Wald entropy of the Schwarzschild black hole in the ghost-free, infinite derivative theory of gravity is
given by the area law~\cite{Conroy:2015wfa}. Such a theory is related to effective nonlocal behaviors of spacetime and thus, would be rather called as super-renormalizable gravity~\cite{Modesto:2011kw}.
In this case, the infrared behavior of gravity is determined by the Einstein-Hilbert action provided that the massless graviton mode remains the only propagating
degree of freedom in the Minkowski spacetime. Its generalization to $D$-dimensional spacetime was considered for deriving the Wald entropy of $D$-dimensional Schwarzschild-Tangherlini black hole~\cite{Conroy:2015nva}.

On the other hand, it is well-known that the Einstein-Gauss-Bonnet (EGB) gravity  which is the zeroth-order approximation of the infinite-derivative gravity provides a constant term to the area law of entropy.
The Gauss-Bonnet term has  an effect on  the semi-classical gravity because it alters the entropy of the black hole. The entropy is not given by the area law but by the Jacobson-Myers entropy formula~\cite{Jacobson:1993xs}.
This is confirmed by three different approaches: the Euclidean action~\cite{Myers:1988ze}, the first-law of thermodynamics $dM=TdS$~\cite{Cai:2001dz},
and  the  Wald entropy formula~\cite{Liu:2008zf,Maeda:2010bu,Sarkar:2010xp}.
In four dimensions, the Gauss-Bonnet term is topological and thus does not affect the equation of motion.
 This implies that the propagating degrees of freedom are  still two of a massless graviton in the EGB gravity.
 Hence, the Wald entropy of the black hole~\cite{Conroy:2015wfa,Conroy:2015nva} computed from  the infinite-derivative theory of gravity
should include   the black hole entropy of the EGB gravity as the zeroth-order approximation. However, it does not work.

In this work, we wish to compute again  the Wald entropy of the Schwarzschild black hole in the infinite-derivative theory of gravity.
Requiring that the massless graviton be the only propagating mode in the Minkowski sapcetime,
the black hole entropy is not  given purely  by the area law, but takes  an additional contribution depending on the power of d'Alembertian operator.

\section{Infinite-derivative theory of gravity}
We start   with  the action for  an infinite-derivative theory of gravity~\cite{Biswas:2011ar}
\begin{eqnarray}
S= S^{\rm EH}+S^{\rm UV}&=&\frac{1}{16\pi G}\int d^4 x\sqrt{-g}\Big\{R +\alpha[R{\cal F}_1(\square)R \nonumber \\
 &+&R_{\mu\nu}{\cal F}_2(\square)R^{\mu\nu}+R_{\mu\nu\rho\sigma}{\cal F}_3(\square)R^{\mu\nu\rho\sigma}]\Big\},\label{Action}
\end{eqnarray}
where $\alpha(>0)$ has inverse of mass squared dimension and  d'Alembertian operator $\square=\frac{1}{\sqrt{-g}}\partial_\mu(\sqrt{-g}g^{\mu\nu}\partial_\nu)$.
$S^{\rm EH}$ represents the Einstein-Hilbert action, while $S^{\rm UV}$ denotes the ultraviolet-modification action.
The ${\cal F}_i$'s are analytic functions of $\square$ given by
\begin{equation} \label{fpow}
{\cal F}_i(\square)=\sum_{n=0}^{\infty} f_{i_n}\Big(\frac{\square}{M^2}\Big)^n,
\end{equation}
where $ f_{i_n}$ are appropriate constants and mass scale  $M \le M_{\rm P}$.
We note that the EGB gravity is recovered from (\ref{Action})  if ${\cal F}_1={\cal F}_3=1$ and ${\cal F}_3=-4$($f_{1_0}=f_{3_0}=1,~f_{2_0}=-4$ and $f_{i_n}=0$ for any $n\ge 1$).
Around Minkowski spacetime, one achieves the bilinear action to (\ref{Action}). In this case,  the ghost-free condition gives a constraint on the form
factors ${\cal F}_i$'s~\cite{Biswas:2013cha}
\begin{equation} \label{ghostf}
2{\cal F}_2(\square)+{\cal F}_2(\square)+2{\cal F}_3(\square)=0.
\end{equation}
This implies that the EGB gravity is the zeroth-order approximation of the ghost-free, infinite derivative gravity.

The full classical equation of motion  has been derived from (\ref{Action})  in~\cite{Biswas:2013cha,Conroy:2014eja} when replacing the Riemann tensor $R_{\mu\nu\rho\sigma}$ by  the Weyl tensor $C_{\mu\nu\rho\sigma}$ in the last term of (\ref{Action}), and Ref.~\cite{Schmidt:1990dh} is referred  for mathematical techniques.
Because  equation of motion takes a lengthly form, we do not want to  write down it here.

One is interested in exploring the connection between the ${\cal F}_i$'s and the gravitational (black hole) entropy.
For this purpose, we  may introduce the Schwarzschild black hole  which is the  solution to Ricci-flat ($R_{\mu\nu}=0$),
\begin{equation} \label{metric}
ds^2_{\rm Sch}=\bar{g}_{\mu\nu} dx^{\mu}dx^{\nu}=\bar{g}_{ab}dx^a dx^b+\bar{g}_{ij}dx^idx^j=-f(r)dt^2+\frac{dr^2}{f(r)}+r^2 d\Omega^2_2
\end{equation}
with the metric function
\begin{equation} \label{sch-sol}
f(r)=1-\frac{2m}{r}\equiv1-\frac{r_H}{r}.
\end{equation}
Here $d\Omega^2_2$ represents the metric of a two-dimensional sphere $S^2$.
At this stage, we would like to mention that if ${\cal F}_3=0$ in (\ref{Action}), all Ricci-flat spacetimes including the Schwarzschild black hole
are exact solutions~\cite{Modesto:2016max}. In case of the EGB gravity, the Schwarzschild spacetime is an exact solution.
If  ${\cal F}_3\not=0$ (non-local), the Schwarzschild spacetime may not be a solution and thus, the contribution to entropy is considered as an off-shell one.
One suggests that the Schwarzschild-de Sitter black hole is the solution to the case of  ${\cal F}_3=0$ in (\ref{Action}). This may be true if one includes the cosmological constant $\Lambda>0$ as
$R-2\Lambda$ in (\ref{Action})~\cite{Conroy:2015nva}. In case of local $F(R)=R+f(R)$ gravity~\cite{Capozziello:2009nq}, one finds
the positive curvature constant $\bar{R}=2f(\bar{R})/(f'(\bar{R})-1)=4\Lambda_f>0$ which may provide the Schwarzschild-de Sitter black hole~\cite{delaCruzDombriz:2009et,Moon:2011hq}. However, the nonlocal case
with $\square$ operators does not allow the constant curvature black hole.  Also, the EGB gravity could not allow the Schwarzschild-de Sitter black hole.

\section{Wald entropy}

 The Wald entropy is defined by the following integral performed on 2-dimensional spacelike bifurcation surface $\Sigma$~\cite{Wald:1993nt,Iyer:1994ys,Jacobson:1993vj}:
 \begin{equation} \label{w-ent}
 S_{\rm W}=-2\pi \oint \Big(\frac{\delta {\cal L}}{\delta R_{\mu\nu\rho\sigma}}\Big)^{(0)} \epsilon_{\mu\nu}\epsilon_{\rho\sigma}dV^2_2,
 \end{equation}
 where $dV^2_2=r^2\sin \theta d\theta d\phi$ is the volume element on $\Sigma$, $\epsilon_{\mu\nu}$ is the binormal vector to $\Sigma$ normalized as $ \epsilon_{\mu\nu} \epsilon^{\mu\nu}=-2$,
 and ${\cal L}=({\cal L}^{\rm EH}+\alpha{\cal L}^{\rm UV})/16\pi G$ is the Lagrangian density in (\ref{Action}).
  The superscript `` (0)" denotes that the functional derivative with respect to $R_{\mu\nu\rho\sigma}$ is evaluated  on shell.
  The variation of the Lagrangian density with respect to  $R_{\mu\nu\rho\sigma}$ is performed as if  $R_{\mu\nu\rho\sigma}$ and the metric $g_{\mu\nu}$ are independent.
 For details, we calculate~\cite{Maeda:2010bu,Bellini:2010ar}
 \begin{eqnarray}
\frac{\delta R}{\delta R_{\mu\nu\rho\sigma}}&=&g^{\mu\rho}g^{\nu\sigma},~~\frac{\delta (R{\cal F}_1R)}{\delta R_{\mu\nu\rho\sigma}}=2g^{\mu\rho}g^{\nu\sigma}{\cal F}_1R,\nonumber \\
\frac{\delta(R_{\alpha\beta}{\cal F}_2R^{\alpha\beta})}{\delta R_{\mu\nu\rho\sigma}}&=&2g^{\mu\rho}{\cal F}_2R^{\nu\sigma},~~\frac{\delta (R_{\alpha\beta\gamma\eta}{\cal F}_3R^{\alpha\beta\gamma\eta})}{\delta R_{\mu\nu\rho\sigma}}=2{\cal F}_3R^{\mu\nu\rho\sigma}.
 \end{eqnarray}
 For the metric (\ref{metric}), $\Sigma$ is given by $t=$ constant and $r=r_H$ and thus, $\epsilon_{tr}=1$.
  Then, one finds
  \begin{eqnarray}
&& \Big(\frac{\delta {\cal L}^{EH}}{\delta R_{\mu\nu\rho\sigma}}\Big) \epsilon_{\mu\nu}\epsilon_{\rho\sigma}=-2, \nonumber \\
 &&\Big( \frac{\delta {\cal L}^{UV}}{\delta R_{\mu\nu\rho\sigma}}\Big) \epsilon_{\mu\nu}\epsilon_{\rho\sigma}=-4{\cal F}_1\bar{R}-2{\cal F}_2~^{(2)}\bar{R}+8{\cal F}_3\bar{R}_{trtr} \nonumber \\
 &&~~~~~~~~~~~~~~~~=-2(2{\cal F}_1+{\cal F}_2+2{\cal F}_3)^{(2)}\bar{R}-4({\cal F}_1\bar{g}^{ij}\bar{R}_{ij}-{\cal F}_3\bar{R}^{ai}~_{ai}),
  \end{eqnarray}
  where the background curvatures are given by~\cite{Dotti:2005rc}
  \begin{eqnarray} \label{curvature}
 \bar{R}&=&\bar{g}^{ab}\bar{R}_{ab}+\bar{g}^{ij}\bar{R}_{ij}=\Big\{-f''-\frac{2f'}{r}\Big\}+\Big\{-\frac{2f'}{r}+\frac{2(1-f)}{r^2}\Big\},\nonumber \\
 ^{(2)}\bar{R}&=&\bar{g}^{ab}\bar{R}_{ab}=\bar{g}^{tt}\bar{R}_{tt}+\bar{g}^{rr}\bar{R}_{rr}=-f''-\frac{2f'}{r},\nonumber \\
 \bar{R}_{trtr}&=&\frac{f''}{2}=-\frac{ ^{(2)}\bar{R}}{2}+\frac{\bar{R}^{ai}~_{ai}}{2},~~\bar{R}^{ai}~_{ai}=-\frac{2f'}{r}.
  \end{eqnarray}
Here the prime ($'$) denotes the differentiation with respect to $r$.

Now, the Wald entropy takes the form
\begin{eqnarray}  \label{entropy1}
S_{\rm W}&=&-\frac{1}{8G}\oint\Big(\frac{\delta {\cal L}}{\delta R_{\mu\nu\rho\sigma}}\Big)^{(0)} \epsilon_{\mu\nu}\epsilon_{\rho\sigma}r^2_H\sin\theta d\theta d\phi \nonumber \\
&=& \frac{A}{4G}\Big[1+\alpha\Big(2{\cal F}_1(\square)+{\cal F}_2(\square)+2{\cal F}_3(\square)\Big)~^{(2)}R\mid_{r\to r_H} \nonumber \\
~~~~~~~~~&+& 2\alpha\Big({\cal F}_1(\square)\bar{g}^{ij}\bar{R}_{ij}\mid_{r\to r_H}-{\cal F}_3(\square)\bar{R}^{ai}~_{ai}\mid_{r\to r_H}\Big)\Big],
\end{eqnarray}
which is exactly the same form as (18) in Ref.~\cite{Conroy:2015nva} with $ A=4\pi r^2_H$.
Imposing the ghost-free condition (\ref{ghostf}), the Wald entropy reduces to
\begin{equation} \label{entropy2}
S_{\rm W}^{\rm ghost-free}=\frac{A}{4G}\Big[1+ 2\alpha\Big({\cal F}_1(\square)\bar{g}^{ij}\bar{R}_{ij}\mid_{r\to r_H}-{\cal F}_3(\square)\bar{R}^{ai}~_{ai}\mid_{r\to r_H}\Big)\Big].
\end{equation}
First of all, we wish to recover the Wald entropy of black hole for  the EGB gravity.
In this case, plugging ${\cal F}_1={\cal F}_3=1$ into (\ref{entropy2})  and using (\ref{curvature}), one finds that
\begin{equation}
\Big(\bar{g}^{ij}\bar{R}_{ij}-\bar{R}^{ai}~_{ai}\Big)\mid_{r\to r_H}=\frac{2(1-f)}{r^2}\mid_{r\to r_H}=\frac{2}{r^2_H}.
\end{equation}
As a result, we obtain the Wald entropy
\begin{equation} \label{entropy3}
S_{\rm W}^{\rm EGB}=\frac{A}{4G}\Big[1+ 4\frac{\alpha}{r^2_H}\Big]=\frac{A}{4G}+\frac{4\alpha\pi}{G},
\end{equation}
which is the exact entropy of the black hole for  the EGB gravity~\cite{Myers:1988ze,Cai:2001dz,Maeda:2010bu}.

For simplicity, we choose the case of ${\cal F}_1(\square)={\cal F}_3(\square)$ which means that $f_{1_i}=f_{3_i}$ for any $i$.
Then, one has
\begin{equation}
{\cal F}_1(\square)(\bar{g}^{ij}\bar{R}_{ij}-\bar{R}^{ai}~_{ai})\mid_{r\to r_H}
=2\sum_{n=0}^{\infty} f_{1_n}\Big(\frac{\square}{M^2}\Big)^n\Big[\frac{1-f}{r^2}\Big]\mid_{r\to r_H},
\end{equation}
which implies that in order to obtain the entropy correctly, one first acts the d'Alembertian operator on $E(r)\equiv(1-f)/r^2=r_H/r^3$  and then, replaces $r$ by $r_H$.
Considering the Schwarzschild metric (\ref{metric}),  the d'Alembertian operator is computed as
\begin{eqnarray} \label{lapalcian}
\square&=&\square_2+\frac{\nabla^2_{(\theta,\phi)}}{r^2}, \nonumber \\
\square_2&=&-\frac{1}{f}\frac{\partial^2}{\partial t^2}+\frac{1}{r^2}\frac{\partial}{\partial r}\Big(r^2f \frac{\partial}{\partial r}\Big), \label{lapla1} \\
\nabla^2_{(\theta,\phi)}&=&\frac{1}{\sin \theta}\frac{\partial}{\partial \theta}\Big(\sin \theta\frac{\partial}{\partial \theta}\Big)+\frac{1}{\sin^2\theta}\frac{\partial^2}{\partial \phi^2}. \nonumber
\end{eqnarray}
Taking into account the radial function of $E(r)$,  $f(r_H)=0$, and $f''(r)=-2f'(r)/r$,  one finds that
\begin{eqnarray}
\square E(r)\mid_{r\to r_H}&=&f'\frac{\partial}{\partial r}E(r)\mid_{r\to r_H}=-\frac{3}{r_H^4}, \nonumber \\
\square^2E(r)\mid_{r\to r_H}&=&\Big[2f'^2\frac{\partial^2}{\partial r^2}+\Big(\frac{2f'^2}{r}+f'f''\Big)\frac{\partial}{\partial r}\Big]E(r)\mid_{r\to r_H}=\frac{12}{r^6_H},\label{dd1} \label{lapla2} \\
\square^3E(r)\mid_{r\to r_H}&=& 4f'^3\Big(\frac{\partial^3}{\partial r^3}-\frac{1}{r^2}\frac{\partial}{\partial r}\Big)E(r)\mid_{r\to r_H} \nonumber \\
                 &+&f'^2\Big(5f''+\frac{6f'}{r}\Big)\frac{\partial^2}{\partial r^2}E(r)\mid_{r\to r_H}=-\frac{12\times23}{r_H^8}.
\end{eqnarray}
At this stage, we compute  up to  $n=3$
\begin{eqnarray}
{\cal F}_1(\square)(\bar{g}^{ij}\bar{R}_{ij}-\bar{R}^{ai}~_{ai})\mid_{r\to r_H}&=&2\Big[\frac{f_{1_0}}{r_H^2}-\frac{3f_{1_1}}{M^2r_H^4}+\frac{12f_{1_2}}{M^4 r_H^6}-\frac{276f_{1_3}}{M^6r_H^8} \nonumber \\
&+&\cdots\Big]. \label{lapla3}
\end{eqnarray}

Finally, the Wald entropy of the black hole in the ghost-free, infinite derivative gravity is given by
\begin{equation} \label{entropy4}
S_{\rm W,{\cal F}_1={\cal F}_3}^{\rm ghostfree}=\frac{A}{4G}\Big[1+4\alpha\Big(\frac{f_{1_0}}{r_H^2}-\frac{3f_{1_1}}{M^2r_H^4}+\frac{12f_{1_2}}{M^4 r_H^6}-\frac{276f_{1_3}}{M^6r_H^8}+\cdots\Big)\Big].
\end{equation}
In  case of $f_{1_0}=1$ and $f_{1_i}=0$ for $i\ge1$, $S_{\rm W,{\cal F}_1={\cal F}_3}^{\rm ghostfree}$ (\ref{entropy4}) leads to the entropy for the EGB gravity $S_{\rm W}^{\rm EGB}$(\ref{entropy3}).
This is an exact result.
We observe here that the entropy depends on the power of  d'Alembertian operator and each operator yields  $1/r_H^2$. Also there exists alternative sign-dependence.  If the mass scale $M$ is chosen to be  comparable  $M_{\rm P}$,
the sum of all higher-order derivative corrections might be  less than $f_{1_0}/r_H^2$ and thus, one expects that there is no the sign of  negative  entropy.
At this stage, we would like to mention that  all higher-order derivative corrections are off-shell contributions to entropy because the Schwarzschild solution (\ref{sch-sol}) might  not be  an exact solution for
action (\ref{Action}).

\section{Discussions}
We have obtained the Wald entropy of the black hole (\ref{entropy4}) for the ghost-free, infinite derivative theory of gravity.
We have recovered the entropy for the  EGB gravity from it when imposing the appropriate constraints. It is worth noting that  (\ref{entropy4}) includes the contribution from the power of d'Alembertian operator.

It was argued that there is a stringent link between the propagating degrees of freedom for the graviton and the gravitational black hole entropy~\cite{Conroy:2015wfa,Conroy:2015nva}.
As long as a higher derivative theory of gravity  does not introduce any extra propagating degrees of freedom in the linearized level and  the IR limit of such a theory
is described by the Einstein gravity, the contribution to the Wald entropy due to the higher-derivative corrections vanishes, yielding the area law of the black hole entropy and thus,
preserving the holographic nature of gravity.  However, this statement does not  seem to be correct.

 A simple counter example is the EGB gravity. Even though its propagating degrees of freedom are  two of  a massless graviton in Minkowski spacetime due to topological nature of Gauss-Bonnet term,
 the entropy is not given by the area law, but  given by (\ref{entropy3}) which shows an additional constant to the area law clearly.
 Another counter example is the infinite-derivative gravity (\ref{Action}). Its propagating degrees of freedom are still  two of a massless graviton if
one chooses  the ghost-free condition (\ref{ghostf}), whereas the Wald entropy  is not given by the area law but it is modified to be   (\ref{entropy4})
 which includes additional corrections depending on the power of d'Alembertian operator  if one requires ${\cal F}_1={\cal F}_3$ further. We note that these additional corrections are off-shell contribution to entropy.

Recently, the gravitational energy-momentum pseudo-tensor $\tau^{\eta}_\alpha$ for higher-order gravity were derived to obtain further modes of gravitational radiation
and to deal with nonlocal theories of gravity~\cite{Capozziello:2017xla}. It could be interesting to see how the entropy of the black hole is related to $\tau^{\eta}_\alpha$.
However, we would like to point out  the difference that $\tau^{\eta}_\alpha(\tilde{\tau}^\eta_\alpha)$ was obtained by varying the Lagrangian with respect to  the metric tensor $g_{\mu\nu}(h_{\mu\nu})$, whereas the Wald entropy (\ref{w-ent}) is obtained from varying the Lagrangian with respect to the curvature tensor $R_{\mu\nu\rho\sigma}$.
It is noted that $\tilde{\tau}^\eta_\alpha$ may be useful to account for  the propagating degrees of freedom for gravitational waves without imposing the ghost-free condition (\ref{ghostf}).

Finally, we wish to comment on possible astrophysical features related to the infinite-derivative gravity.
One suggests that  the black hole shadow might be related to the further terms emerging from (\ref{entropy4}).
Let us explain briefly  what is the black hole shadow. Part of the photons emitted from a luminous background behind a black hole  end up falling into black hole, not reaching the observer,
and producing a completely dark zone denominated the shadow. The size and the shape of the shadow depend on the mass and the angular momentum of the black hole.
The apparent shape of a black hole is thus defined by the boundary of the shadow. The shadow of the Schwarzschild black hole is circular,
 but the presence of spin produces a deformation~\cite{Vazquez:2003zm,Hioki:2009na}. The analysis of the shadow will be a useful tool for extracting properties of astrophysical black holes and 
 comparing gravitational theories. The Event Horizon Telescope will observe the shadow of the supermassive Galactic black hole soon~\cite{Johannsen:2015mdd}.
The boundary of the shadow may be connected  to the black hole entropy  because the entropy is proportional to the area of  the  astrophysical black hole.
In this sense,  the black hole shadow might be related to the further terms emerging from (\ref{entropy4}) in the infinite-derivative gravity.

\section*{Acknowledgement}
The author thanks to Leonardo Modesto for helpful discussions.
This work was supported by the National Research Foundation of Korea (NRF) grant funded by the Korea government (MOE) (No. NRF-2017R1A2B4002057).

\end{document}